\documentclass[journal,10pt]{IEEEtran}

\usepackage{amsmath,amssymb,amsthm}

\usepackage{graphicx}
\usepackage[caption=false,font=footnotesize]{subfig}
\usepackage{epstopdf}
\usepackage{booktabs,threeparttable}
\usepackage{array,mdwmath,mdwtab,stfloats}
\usepackage{multicol}
\usepackage{amsmath}

\usepackage{algorithm}
\usepackage{algpseudocode}
\algrenewcommand{\algorithmicrequire}{\textbf{Input:}}
\algrenewcommand{\algorithmicensure}{\textbf{Output:}}

\usepackage{url}
\usepackage{xcolor}

\usepackage[colorlinks]{hyperref}

\newcommand{\MeijerG}[7]{G_{#1,#2}^{#3,#4} \left( #5 \middle| \begin{smallmatrix} #6 \\ #7 \end{smallmatrix} \right)}

\newcommand{\SDS}{\mathbb{S}_{DS}}

\begin{document}
\title{\LARGE Fluid Antenna System-Enabled UAV-to-Ground  Communications}

\author{Xusheng Zhu, 
            Kai-Kit Wong, \IEEEmembership{Fellow, IEEE},
            Qingqing Wu,
            Hyundong Shin, \IEEEmembership{Fellow, IEEE}, and
            Yangyang Zhang
\vspace{-9mm}

\thanks{(\emph{Corresponding author: Kai-Kit Wong}.)}
\thanks{X. Zhu and K. K. Wong are with Department of Electronic and Electrical Engineering, University College London, London, United Kingdom (e-mail: \{xusheng.zhu,kai-kit.wong\}@ucl.ac.uk). K. K. Wong is also affiliated with the Department of Electronic Engineering, Kyung Hee University, Yongin-si, Gyeonggi-do 17104, Republic of Korea.}
\thanks{Q. Wu is with the Department of Electronic Engineering, Shanghai Jiao Tong University, Shanghai 200240, China (e-mail: qingqingwu@sjtu.edu.cn).}
\thanks{H. Shin is with the Department of Electronics and Information Convergence Engineering, Kyung Hee University, Yongin-si, Gyeonggi-do 17104, Republic of Korea (e-mail: hshin@khu.ac.kr).}
\thanks{Y. Zhang is with Kuang-Chi Science Limited, Hong Kong SAR, China (e-mail: yangyang.zhang@kuang-chi.org).}
}

\maketitle

\begin{abstract}
Fluid antenna systems (FAS) have emerged as a revolutionary technology offering enhanced spatial diversity within a compact form factor. Concurrently, unmanned aerial vehicles (UAVs) are integral to future networks, necessitating channel models that capture both multipath fading and shadowing. This letter presents a novel performance analysis of a UAV-to-ground link, where the receiver is equipped with an $N$-port FAS operating over the challenging double-shadowing fading channel. By adapting a tractable eigenvalue-based approximation for the correlated FAS ports, we derive new analytical expressions for the end-to-end signal-to-noise ratio statistics, namely the cumulative distribution function and the probability density function. Based on these statistics, we present exact integral expressions for the outage probability, average bit error rate, and average channel capacity. We further derive new, tractable closed-form solutions for the average bit error rate and capacity for the practical dual-rank, independent but non-identically distributed case. Finally, a key asymptotic analysis reveals that the system achieves a multiplicative diversity order of $G_d = M \times d$, which is precisely the product of the FAS spatial rank $M$ and the intrinsic channel diversity order $d$. Simulation results are provided to validate the high accuracy of our entire theoretical framework.
\end{abstract}


\IEEEpeerreviewmaketitle

\section{Introduction}\label{sec:introduction}
\IEEEPARstart{U}{nmanned} aerial vehicles (UAVs) are poised to become a critical component of future wireless networks, offering flexible, on-demand deployment for applications ranging from aerial base stations (BSs) to mobile relays \cite{jiang2025dyn}. But the unique characteristics of the air-to-ground (A2G) channel, which differs greatly from terrestrial models, presents a major communication challenge \cite{polus2023uav}. The A2G link is often subject to both small-scale multipath fading and large-scale shadowing from buildings, foliage, or the UAV airframe itself \cite{bithas2020uav}. To capture this behavior accurately, the double-shadowed (DS) fading model is widely adopted, which composites Nakagami-$m$ fading with Inverse-Gamma (IG) shadowing, resulting in a tractable Meijer's G-function representation \cite{li2024rad}.

To strengthen the A2G channel, the emerging fluid antenna system (FAS) concept seems useful. FAS treats antenna as a reconfigurable physical-layer resource to empower the physical layer with shape and position flexibility \cite{wong2020pel,wong2021fa,zhu2025fris}, and has recently attracted much interest \cite{new2025a,Lu-2025,New-2026jsac}. FAS is especially appealing in space-constrained devices by allowing one radio-frequency (RF) chain to effectively retrieve spatial diversity over a compact aperture \cite{zhu2025fas,zhu2023scal}. While FAS is a natural fit for space-constrained UAV ground receivers, the performance of an FAS-enabled receiver operating over the complex, non-Rayleigh DS channel remains unknown.

Technically speaking, the correlation among the FAS ports makes direct analysis of the maximum signal-to-noise ratio (SNR) extremely challenging. In \cite{zhao2025fas}, a tractable eigenvalue-based framework was proposed, which models the $N$ correlated ports as $M$ independent but non-identically distributed (i.n.i.d.) eigen-modes ($M \le N$), and is shown to be highly accurate for Rayleigh channels. However, the required analysis for FAS-UAV is highly non-trivial, requiring the unification of the complex eigenvalue-based FAS approximation with the Meijer's G-function statistics of the DS channel.

In this letter, we bridge this significant gap. First, we derive the foundational end-to-end statistics, i.e., cumulative density function (CDF)/probability density function (PDF), for an $M$-rank FAS receiver over the DS channel by successfully integrating the eigenvalue approximation with the G-function-based channel statistics. Second, based on this new framework, we derive exact analytical expressions for the system's outage probability, average bit error rate (ABER), and average channel capacity. Third, we derive new tractable closed-form solutions for the ABER and average capacity for the $M=2$ i.n.i.d.~case. Finally, we carry out an asymptotic analysis at high SNR, proving that the system achieves a total diversity order of $G_d = M \times d$. This key result reveals the overall diversity to be the \textit{product} of the FAS spatial rank $M$ and the intrinsic channel diversity order $d$. Monte-Carlo simulations validate the high accuracy of our framework against both the approximation model and the exact $N$-port physical simulation.

\section{System and Channel Models}\label{sec:system_model}
\begin{figure}
\centering
\includegraphics[width=0.6\columnwidth]{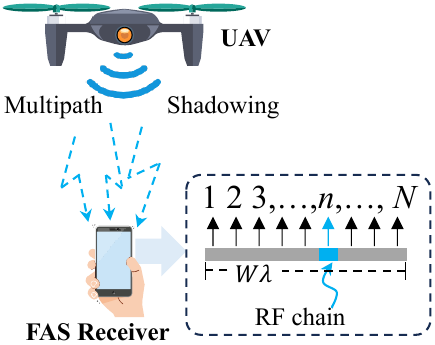}
\caption{The UAV-G link with an $N$-port FAS receiver.}\label{fig:system}
\vspace{-2mm}
\end{figure}

As shown in Fig.~\ref{fig:system}, we consider a point-to-point communication system where a single-antenna transmitter (Tx) mounted on a UAV communicates with a ground-based receiver (Rx). The Rx is equipped with an $N$-port FAS, where the $N$ ports are assumed to be equally spaced along a linear dimension of length $W\lambda$, with $\lambda$ being the carrier wavelength. These $N$ ports are connected to a single RF chain, and the receiver selects the port that achieves the maximum SNR.

\subsection{UAV-to-Ground (UAV-G) Channel Model}\label{subsec:uav_channel}
We adopt the statistical fading channel model developed for UAV-G communications, which accurately accounts for both multipath fading and shadowing effects. We focus on the DS model as it represents the more general and challenging scenario where shadowing is present in the vicinity.

The DS model assumes the channel gain is a product of four independent random variables: $h = X_1 X_2 Y_1 Y_2$, in which $X_j$ models IG shadowing and $Y_j$ models Nakagami-$m$ multipath fading, for $j \in \{1, 2\}$. The instantaneous received power is $P_r \propto |h|^2$, and thus the instantaneous SNR $\gamma = P_r / N_0$ is a product of four independent power terms. Specifically, as $X_j^2$ follows an IG distribution, its PDF is given by
\begin{equation}
f_{X_j^2}(x) = \frac{(\alpha_j-1)^{\alpha_j}}{\Gamma(\alpha_j)} x^{-\alpha_j-1} e^{-\frac{\alpha_j-1}{x}},~x > 0,
\end{equation}
where $\Gamma(\cdot)$ is the Gamma function. Also, as $Y_j^2$ is distributed according to a Gamma distribution, the corresponding PDF is expressed as
\begin{equation}
f_{Y_j^2}(y) = \frac{m_j^{m_j}}{\Gamma(m_j)} y^{m_j-1} e^{-m_j y},~y > 0.
\end{equation}
The PDF of the product of these four random variables (RVs), $\gamma$, can be found using the Mellin transform, which results in the Meijer G-function representation given by \cite[(4)]{polus2023uav} as
\begin{equation}\label{eq:pdf_ds}
f_{\gamma_{\rm DS}}(\gamma) = \gamma^{-1} \SDS \MeijerG{2}{2}{2}{2}{\frac{m_1 m_2}{\overline{\gamma}}\gamma}{1-\alpha_2, 1-\alpha_1}{m_1, m_2},
\end{equation}
where $\overline{\gamma} = \mathbb{E}[\gamma]$ denotes the average SNR, and $m_j$ and $\alpha_j$ are the shape parameters, and $G_{p,q}^{m,n}(\cdot)$ denotes the Meijer's G-function \cite{gradshteyn2014integrals}. Additionally, the constant $\SDS$ is defined as $\SDS = \left[ \Gamma(m_1)\Gamma(m_2)\Gamma(\alpha_1)\Gamma(\alpha_2) \right]^{-1}$. Therefore, the corresponding CDF of the SNR is derived by integrating the PDF, $F_{\gamma_{DS}}(\gamma) = \int_0^\gamma f_{\gamma_{DS}}(t) dt$, which yields
\begin{equation}\label{eq:cdf_ds}
F_{\gamma_{\rm DS}}(\gamma) = \SDS \MeijerG{3}{3}{2}{3}{\frac{m_1 m_2}{\overline{\gamma}}\gamma}{1-\alpha_2, 1-\alpha_1, 1}{m_1, m_2, 0}.
\end{equation}

\subsection{FAS Receiver Model}\label{subsec:fas_model}
We assume that the $N$-port FAS at the receiver always selects the port with the highest instantaneous SNR. The output SNR of the FAS is therefore given by
\begin{equation}\label{eq:fas_max}
\gamma_{\rm FAS} = \max\{\gamma_1, \gamma_2,\dots, \gamma_N\},
\end{equation}
where $\gamma_n = \overline{\gamma}|h_n|^2$ is the SNR at the $n$-th port. Due to the compact physical placement of the $N$ ports, the channel gains $\{h_n\}_{n=1}^N$ are spatially correlated. This spatial correlation is commonly captured by the Jakes' model, which defines the $(p,q)$-th entry of the covariance matrix $\mathbf{J}$ as \cite{New2023fluid,new2023information}
\begin{equation}\label{eq:jakes_model}
[\mathbf{J}]_{p,q} = J_0\left(2\pi \frac{|p-q|}{N-1} W\right),
\end{equation}
where $p,q \in \{1,\dots, N\}$ and $J_0(\cdot)$ is the zeroth-order Bessel function of the first kind. The analysis of the maximum of these $N$ correlated RVs is generally intractable.

To overcome this, we adopt the novel analytical framework proposed in \cite{zhao2025fas}. This approach approximates the statistics of the maximum of $N$ correlated RVs by the statistics of the maximum of $M$ i.n.i.d. RVs. Here, $M = {\rm Rank}\{\mathbf{J}\}$ is the number of non-zero eigenvalues of the channel covariance matrix $\mathbf{J} = \mathbb{E}\{\mathbf{h}\mathbf{h}^H\}$, where $\mathbf{h} = [h_1,\dots, h_N]^T$. By performing an eigenvalue decomposition on ${\bf J}$, we have
\begin{equation}
\mathbf{J} = \mathbf{U} \mathbf{\Lambda} \mathbf{U}^H,
\end{equation}
where $\mathbf{U}$ is a unitary matrix containing the eigenvectors and $\mathbf{\Lambda} = {\rm diag}(\lambda_1, \dots, \lambda_M, 0, \dots, 0)$ is the diagonal matrix of the $M$ non-zero eigenvalues, $\lambda_i$. The rank $M \le N$ represents the effective spatial degrees of freedom. This decomposition allows the channel vector $\mathbf{h}$ to be represented as $\mathbf{h} = \mathbf{U} \mathbf{\Lambda}^{1/2} \mathbf{z}$, where $\mathbf{z}$ is a vector of independent and identically distributed (i.i.d.) complex Gaussian RVs. The approximation framework leverages this by modeling the system as $M$ independent branches, each weighted by eigenvalue $\lambda_i$.

We generalize this approximation from \cite[(20)]{zhao2025fas} to our non-Rayleigh DS shadowed fading channels. The output SNR of the FAS is approximated as
\begin{equation}\label{eq:fas_approx}
\gamma_{\rm FAS} \approx \gamma_{\rm approx} = \max\left\{\lambda_1 \gamma^{(1)}, \lambda_2 \gamma^{(2)},\dots, \lambda_M \gamma^{(M)}\right\},
\end{equation}
where $\{\gamma^{(i)}\}_{i=1}^M$ are i.i.d.~RVs, each following $F_{\gamma_{\rm DS}}(\cdot)$. 

\subsection{Statistics of the FAS Output SNR}\label{subsec:fas_stats}
Based on the approximation \eqref{eq:fas_approx}, we now derive the novel statistical expressions for the FAS output SNR, $\gamma_{\rm FAS}$.

\subsubsection{CDF}
The CDF of $\gamma_{\rm FAS}$ is defined as $F_{\gamma_{\rm FAS}}(\gamma) = \mathbb{P}(\gamma_{\rm FAS} \le \gamma)$. Using \eqref{eq:fas_approx}, the CDF is expressed as
\begin{IEEEeqnarray}{rCl}
F_{\gamma_{\rm FAS}}(\gamma) &=& \mathbb{P}\left(\max_{\{i=1,\ldots,M\} }\{\lambda_i \gamma^{(i)}\} \le \gamma\right) \nonumber \\
&\stackrel{(a)}{=}& \prod_{i=1}^{M} \mathbb{P}(\lambda_i \gamma^{(i)} \le \gamma) = \prod_{i=1}^{M} \mathbb{P}(\gamma^{(i)} \le \gamma / \lambda_i) \nonumber \\
&=& \prod_{i=1}^{M} F_{\gamma_{\rm DS}}(\gamma / \lambda_i),\label{eq:cdf_general}
\end{IEEEeqnarray}
where step $(a)$ follows from the independence of the $\gamma^{(i)}$ RVs, and $F_{\gamma_{\rm DS}}(\cdot)$ is the single-link DS CDF from \eqref{eq:cdf_ds}. Substituting \eqref{eq:cdf_ds} into \eqref{eq:cdf_general} yields the new CDF
\begin{equation}\label{eq:cdf_fas_ds}
F_{\gamma_{\rm FAS}}(\gamma) = \prod_{i=1}^{M} \left[ \SDS \MeijerG{3}{3}{2}{3}{\frac{m_1 m_2}{\overline{\gamma}\lambda_i}\gamma}{1-\alpha_2, 1-\alpha_1, 1}{m_1, m_2, 0} \right].
\end{equation}
This is a new and tractable expression for the end-to-end SNR of the FAS-UAV system.

\subsubsection{PDF}
The PDF is obtained by differentiating the CDF, $f_{\gamma_{\rm FAS}}(\gamma) = \frac{d}{d\gamma} F_{\gamma_{\rm FAS}}(\gamma)$. Applying the product rule for differentiation to \eqref{eq:cdf_general} gives
\begin{equation}\label{eq:pdf_general}
f_{\gamma_{\rm FAS}}(\gamma)=\sum_{j=1}^{M} \left[ \left( \frac{1}{\lambda_j} f_{\gamma_{DS}}(\gamma / \lambda_j) \right) \prod_{i \ne j}^{M} F_{\gamma_{DS}}(\gamma / \lambda_i) \right].
\end{equation}
Substituting the single-link PDF \eqref{eq:pdf_ds} and CDF \eqref{eq:cdf_ds} into \eqref{eq:pdf_general} yields the final PDF
\begin{multline}\label{eq:pdf_fas_ds}
f_{\gamma_{\rm FAS}}(\gamma)
=\frac{(\SDS)^M}{\gamma} \sum_{j=1}^{M} \left[ \MeijerG{2}{2}{2}{2}{\frac{m_1 m_2}{\overline{\gamma}\lambda_j}\gamma}{1-\alpha_2, 1-\alpha_1}{m_1, m_2} \right.\\
\left. \times \prod_{i \ne j}^{M} \MeijerG{3}{3}{2}{3}{\frac{m_1 m_2}{\overline{\gamma}\lambda_i}\gamma}{1-\alpha_2, 1-\alpha_1, 1}{m_1, m_2, 0} \right].
\end{multline}

\section{Performance Analysis}\label{sec:performance_analysis}
We now leverage the new statistical characterizations to evaluate the primary performance metrics of the system.

\subsection{Outage Probability}
\label{subsec:pout}
The outage probability is the probability that the FAS output SNR, $\gamma_{\rm FAS}$, falls below a predefined threshold $\gamma_{\rm th}$. This is formally defined by integrating the PDF as
\begin{equation}\label{eq:pout_def}
P_{\rm out} = \int_0^{\gamma_{\rm th}} f_{\gamma_{\rm FAS}}(\gamma) d\gamma = F_{\gamma_{\rm FAS}}(\gamma_{\rm th}).
\end{equation}
By substituting $\gamma_{\rm th}$ directly into our new CDF expression \eqref{eq:cdf_fas_ds}, we obtain the novel outage probability in a tractable form as
\begin{equation}\label{eq:pout_ds}
P_{\rm out} = \prod_{i=1}^{M} \left[ \SDS \MeijerG{3}{3}{2}{3}{\frac{m_1 m_2}{\overline{\gamma}\lambda_i}\gamma_{\rm th}}{1-\alpha_2, 1-\alpha_1, 1}{m_1, m_2, 0} \right].
\end{equation}
For the practical dual-rank case (i.e., $M=2$) with distinct eigenvalues $\lambda_1$ and $\lambda_2$, this expression expands to
\begin{equation}\label{eq:pout_ds_m2}
P_{\rm out}^{M=2} = (\SDS)^2 G_{3,3}^{2,3}\left(c_1\gamma_{\rm th}\Big|\begin{smallmatrix} \mathbf{a} \\ \mathbf{b} \end{smallmatrix}\right) G_{3,3}^{2,3}\left(c_2\gamma_{\rm th}\Big|\begin{smallmatrix} \mathbf{a} \\ \mathbf{b} \end{smallmatrix}\right),
\end{equation}
where $c_i = m_1 m_2/(\overline{\gamma}\lambda_i)$, $\mathbf{a}=\{1-\alpha_2, 1-\alpha_1, 1\}$, and $\mathbf{b}=\{m_1, m_2, 0\}$.

\subsection{ABER}
\label{subsec:aber}
The ABER is calculated by averaging the conditional error probability (CEP) over the SNR distribution as
\begin{equation}
P_{\rm ABER} = \int_{0}^{\infty} P_e(\gamma) f_{\gamma_{\rm FAS}}(\gamma) d\gamma,
\end{equation}
where $P_e(\gamma)$ is the CEP for a given modulation. For binary phase shift keying (BPSK), $P_e(\gamma) = Q(\sqrt{2\kappa\gamma})$, where $Q(\cdot)$ is the Gaussian Q-function and $\kappa=1$. Using the alternative form $Q(x) = \frac{1}{\pi} \int_0^{\pi/2} \exp(-\frac{x^2}{2\sin^2\theta}) d\theta$, the ABER is found as
\begin{align}
P_{\rm ABER} &= \frac{1}{\pi} \int_{0}^{\pi/2} \mathbb{E}\left[e^{-\frac{\kappa\gamma_{\rm FAS}}{\sin^2\theta}}\right] d\theta\notag\\
&= \frac{1}{\pi} \int_{0}^{\pi/2} \mathcal{M}_{\gamma_{\rm FAS}}\left(\frac{\kappa}{\sin^2\theta}\right) d\theta,
\end{align}
where $\mathcal{M}_{\gamma_{\rm FAS}}(s) = \mathbb{E}[e^{-s\gamma_{\rm FAS}}]$ is the moment generating function (MGF) of $\gamma_{\rm FAS}$. However, the MGF of the maximum of i.n.i.d.~G-distributed RVs is intractable. Note that a more tractable approach, which utilizes the CDF, is given by \cite[(20)]{polus2023uav} for several modulation schemes as
\begin{equation}\label{eq:aber_def}
P_{\rm ABER} = \frac{\kappa^{\beta}}{2\Gamma(\beta)}\int_{0}^{\infty}\gamma^{\beta-1} e^{-\kappa\gamma} F_{\gamma_{\rm FAS}}(\gamma) d\gamma,
\end{equation}
where $(\kappa, \beta)=(1, 0.5)$ for BPSK. Substituting our new CDF $F_{\gamma_{\rm FAS}}(\gamma)$ from \eqref{eq:cdf_fas_ds} into \eqref{eq:aber_def} and applying the Taylor expansion $e^{-\kappa\gamma} = \sum_{n=0}^{\infty} \frac{(-\kappa)^n}{n!}$, yields the series form
\begin{equation}\label{eq:aber_series}
P_{\rm ABER} = \frac{\kappa^{\beta}}{2\Gamma(\beta)} \sum_{n=0}^{\infty} \frac{(-\kappa)^n}{n!} \mathcal{I}_n,
\end{equation}
where $\mathcal{I}_n$ is the $n$-th moment integral
\begin{equation}\label{eq:aber_integral}
\mathcal{I}_n = \int_{0}^{\infty}\gamma^{\beta+n-1} F_{\gamma_{\rm FAS}}(\gamma) d\gamma.
\end{equation}
For the practical dual-rank case ($M=2$), $\mathcal{I}_n$ becomes
\begin{multline}\label{eq:aber_int_m2}
\mathcal{I}_n = \int_{0}^{\infty} \gamma^{\beta+n-1} \left[ \SDS \MeijerG{3}{3}{2}{3}{c_1\gamma}{\mathbf{a}}{\mathbf{b}} \right]\\
\times \left[ \SDS \MeijerG{3}{3}{2}{3}{c_2\gamma}{\mathbf{a}}{\mathbf{b}} \right] d\gamma,
\end{multline}
where $c_i = m_1 m_2 / (\overline{\gamma}\lambda_i)$, $\mathbf{a}=\{1-\alpha_2, 1-\alpha_1, 1\}$, and $\mathbf{b}=\{m_1, m_2, 0\}$. This integral can be solved in closed form using \cite[(7.811)]{gradshteyn2014integrals}. For the $M=2$ i.n.i.d.~case ($c_1 \ne c_2$), the integral evaluates to
\begin{equation}\label{eq:aber_sol_m2}
\mathcal{I}_n= (\SDS)^2 \left(\frac{1}{c_1}\right)^{\beta+n} \MeijerG{6}{6}{5}{5}{\frac{c_2}{c_1}}{ 1-\mathbf{b}-(\beta+n), \mathbf{a} }{ \mathbf{a}-(\beta+n), \mathbf{b} },
\end{equation}
where the vector notation $\mathbf{a}-(\beta+n)$ implies subtraction from the first two elements only, i.e., $\{1-\alpha_2-\beta-n, 1-\alpha_1-\beta-n, 1\}$, and $1-\mathbf{b}-(\beta+n)$ is $\{1-(m_1+\beta+n), 1-(m_2+\beta+n), 1-(\beta+n)\}$. Substituting \eqref{eq:aber_sol_m2} into \eqref{eq:aber_series} provides a new, exact analytical expression for the ABER.

\begin{figure*}[t]
\centering
\subfloat[Outage probability vs. average SNR.]{%
\includegraphics[width=0.31\textwidth]{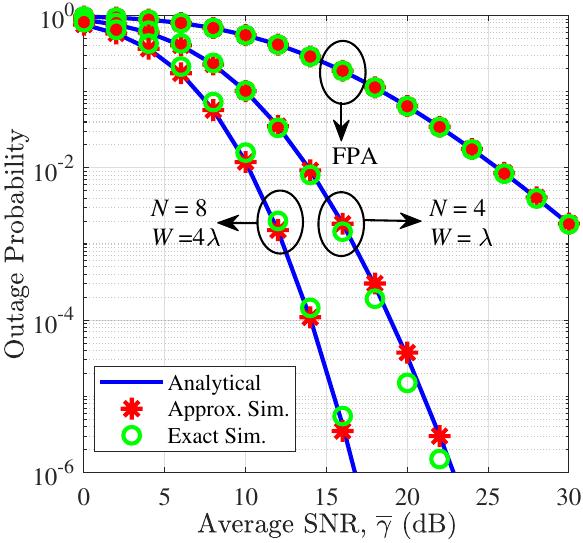}\label{fig:fig1a}%
}\hfill
\subfloat[ABER vs. average SNR.]{%
\includegraphics[width=0.31\textwidth]{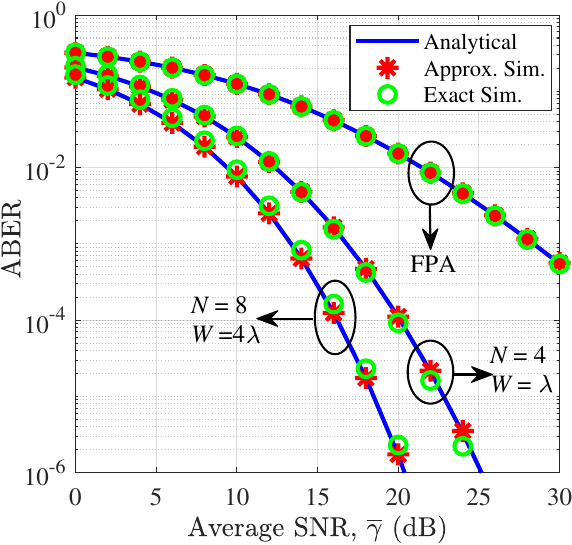}\label{fig:fig1b}%
}\hfill
\subfloat[Average channel capacity vs. average SNR.]{%
\includegraphics[width=0.31\textwidth]{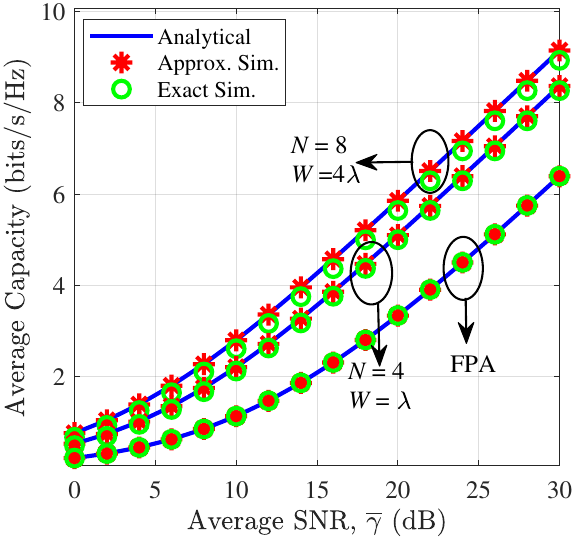}\label{fig:fig1c}%
}
\caption{Analytical results validation and FAS diversity gain ($M=N$).}\label{fig:fig1_full_validation}
\end{figure*}

\subsection{Average Channel Capacity}\label{subsec:capacity}
The average channel capacity, $\overline{C}$, is defined by
\begin{equation}\label{eq:cap_def_pdf}
\overline{C} = \mathbb{E}[B \log_2(1+\gamma_{\rm FAS})] = \frac{B}{\ln(2)} \int_{0}^{\infty} \ln(1+\gamma) f_{\gamma_{\rm FAS}}(\gamma) d\gamma.
\end{equation}
Using integration by parts, $\overline{C} = \frac{B}{\ln(2)} [ \ln(1+\gamma) F_{\gamma_{\rm FAS}}(\gamma) ]_0^\infty - \int_0^\infty \frac{F_{\gamma_{\rm FAS}}(\gamma)}{1+\gamma} d\gamma$, which simplifies to the numerically stable form
\begin{equation}\label{eq:cap_def_cdf}
\overline{C} = \frac{B}{\ln(2)} \int_{0}^{\infty} \frac{1 - F_{\gamma_{\rm FAS}}(\gamma)}{1+\gamma} d\gamma.
\end{equation}
While \eqref{eq:cap_def_cdf} is convenient for numerical integration, solving \eqref{eq:cap_def_pdf} allows for a closed-form solution for $M=2$. Substituting our PDF \eqref{eq:pdf_general} into \eqref{eq:cap_def_pdf} gives
\begin{multline}\label{eq:cap_int_general}
\overline{C} =\frac{B}{\ln(2)} \sum_{j=1}^{M} \int_{0}^{\infty} \ln(1+\gamma) \left[ \frac{1}{\lambda_j} f_{\gamma_{DS}}(\gamma / \lambda_j) \right. \notag\\
 \left. \times \prod_{i \ne j}^{M} F_{\gamma_{DS}}(\gamma / \lambda_i) \right] d\gamma.
\end{multline}
For the $M=2$ i.n.i.d.~case, we solve this by representing the logarithm term as a Meijer G-function \cite{gradshteyn2014integrals}
\begin{equation}\label{eq:log_g_func}
\ln(1+\gamma) = \MeijerG{2}{2}{1}{2}{\gamma}{1, 1}{1, 0}.
\end{equation}
The capacity is then given by
\begin{equation}\label{eq:cap_sol_m2}
\overline{C}^{M=2} = \frac{B(\SDS)^2}{\ln(2)} \left[ \mathcal{J}(c_1, c_2) + \mathcal{J}(c_2, c_1) \right],
\end{equation}
where $c_i = m_1 m_2/(\overline{\gamma}\lambda_i)$ and $\mathcal{J}(c_1, c_2)$ is a Bivariate Meijer's G-function that solves the integral of the three G-functions (logarithm, PDF, and CDF). This integral generalizes the i.i.d.~analysis in \cite{polus2023uav} and is given by
\begin{IEEEeqnarray}{rCl}
    \mathcal{J}(c_1, c_2) &=& \int_{0}^{\infty} \MeijerG{2}{2}{1}{2}{\gamma}{1, 1}{1, 0} \left( \gamma^{-1} \MeijerG{2}{2}{2}{2}{c_1\gamma}{\mathbf{c}}{\mathbf{d}} \right) \nonumber \\
    && \times \left[ \MeijerG{3}{3}{2}{3}{c_2\gamma}{\mathbf{a}}{\mathbf{b}} \right] d\gamma \nonumber \\
    &=& G_{2,2:3,3:2,2}^{2,1:2,3:2,2}
    \left[
        \begin{array}{@{}c@{}} 0, 1 \\ 0, 0 \end{array}
        \middle|
        \begin{smallmatrix} \mathbf{a} \\ \mathbf{b} \end{smallmatrix}
        \middle|
        \begin{smallmatrix} \mathbf{c} \\ \mathbf{d} \end{smallmatrix}
    \middle| c_2, c_1
    \right],
    \label{eq:bivariate_g}
\end{IEEEeqnarray}
with $\mathbf{a}=\{1-\alpha_2, 1-\alpha_1, 1\}$, $\mathbf{b}=\{m_1, m_2, 0\}$, $\mathbf{c}=\{1-\alpha_2, 1-\alpha_1\}$, and $\mathbf{d}=\{m_1, m_2\}$.

\subsection{Asymptotic Analysis: Diversity Order}\label{subsec:diversity}
To gain simpler insights, we analyze the outage probability in the high-SNR regime where $\overline{\gamma} \to \infty$. First, we require the asymptotic behavior of the single-link CDF from \eqref{eq:cdf_ds}. For cascaded fading channels, the CDF at $\gamma \to 0$ behaves as $F_{\gamma_{UAV}}(\gamma) \propto \gamma^d$, where $d$ is the single-link diversity order, given by $d = \min(\alpha_1, \alpha_2, m_1, m_2)$ for the DS case \cite[(24)]{polus2023uav}.
Based on this, we analyze the CDF from \eqref{eq:cdf_general} as
\begin{equation}
\begin{aligned}
    P_{\rm out} &= F_{\gamma_{\rm FAS}}(\gamma_{\rm th}) = \prod_{i=1}^{M} F_{\gamma_{\rm DS}}(\gamma_{\rm th} / \lambda_i) \nonumber \\
    &\approx \prod_{i=1}^{M} \left( K \cdot (\gamma_{\rm th} / \lambda_i)^d \right) = K^M (\gamma_{\rm th})^{Md} \left( \prod_{i=1}^M \lambda_i \right)^{-d}.
    \label{eq:pout_asymp}
\end{aligned}
\end{equation}
The diversity orde is defined as $G_d = -\lim\limits_{\overline{\gamma} \to \infty} \frac{\log P_{\rm out}}{\log \overline{\gamma}}$. Since $K$, $\gamma_{th}$, and $\lambda_i$ are constants with respect to $\overline{\gamma}$, but the constant $K$ itself contains $\overline{\gamma}^{-d}$ (from the normalized G-function argument), $P_{\rm out} \propto (\overline{\gamma}^{-d})^M = (\overline{\gamma})^{-Md}$. Thus, we find the system diversity order $G_d$ to be
\begin{equation}\label{eq:diversity_order}
G_d = M \times d.
\end{equation}
This is a powerful and intuitive result. It proves that the total system diversity order is the product of the spatial diversity rank $M$ afforded by the FAS \cite{zhao2025fas} and the intrinsic diversity order $d$ of the single-link shadowed fading channel.

\section{Numerical Results}
In this section, we present numerical and simulation results to validate the accuracy of our derived analytical framework and to quantify the performance of the proposed FAS-enabled UAV-G link over DS channels. Unless stated otherwise, the number of independent realizations is $2\times10^6$. The baseline DS channel parameters are adopted as $m_1=2.1, m_2=2.4, \alpha_1=2.1$, and $\alpha_2=2.4$. The outage threshold is set to $\gamma_{\rm th} = 0$ dB, and the ABER is evaluated for BPSK ($\kappa=1, \beta=0.5$).

Fig.~\ref{fig:fig1_full_validation} presents the validation for the three primary performance metrics. For all metrics and all FAS configurations, the analytical curves, the approximate simulation, and the exact physical simulation are in perfect agreement. This result is highly significant, as it validates the two foundational pillars of this work: 1) first, the eigenvalue-based approximation from \cite{zhao2025fas} is robust and highly accurate even when applied to the complex non-Rayleigh DS channel, and also 2) our novel, G-function-based analytical expressions are valid. Additionally, Figs.~\ref{fig:fig1a} and \ref{fig:fig1b} clearly illustrate the substantial diversity gain provided by FAS. The slope of the outage probability/ABER curves becomes progressively steeper as the effective rank $M$ increases. The fixed-position antenna (FPA) case exhibits a diversity order limited by the channel parameters, $d = 2.1$. The $M=4$ and $M=8$ configurations show significantly steeper slopes, visually confirming our asymptotic diversity order analysis in \eqref{eq:diversity_order}, where the total diversity order $G_d = M \times d$. Fig.~\ref{fig:fig1c} further confirms the multiplexing gain, showing that a higher rank $M$ provides a greater capacity slope.

\begin{figure}[!t]
\centering
\subfloat[Impact of multipath fading ($m$).]{%
\includegraphics[width=0.47\columnwidth]{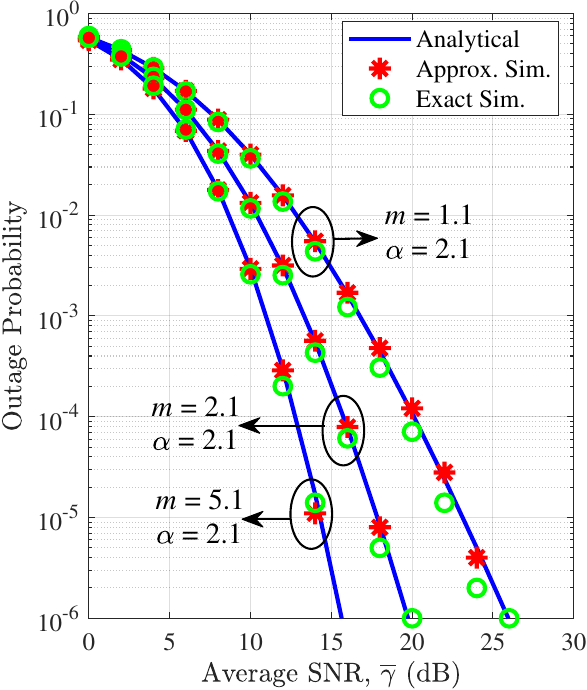}\label{fig:fig2a}%
}
\hfil
\subfloat[Impact of shadowing ($\alpha$).]{%
\includegraphics[width=0.47\columnwidth]{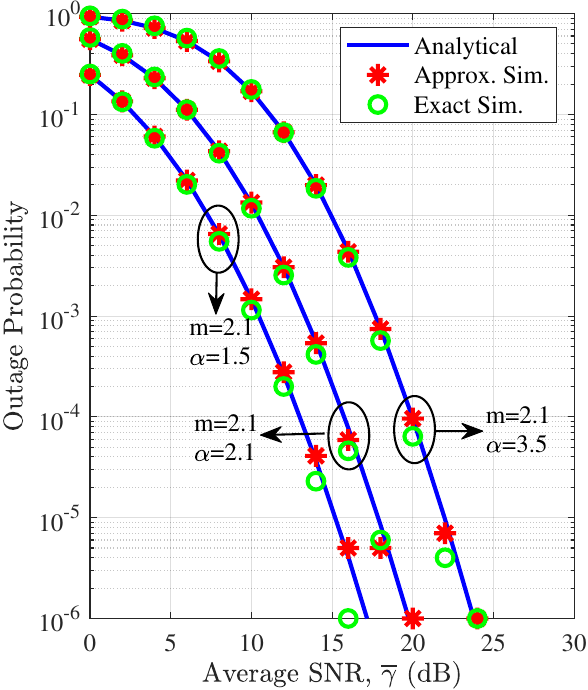}\label{fig:fig2b}%
}
\caption{Impact of UAV parameters on outage probability ($N=4, W=1\lambda, M = 4$).}\label{fig:fig2_uav_params}
\vspace{-2mm}
\end{figure}

Fig.~\ref{fig:fig2_uav_params} investigates the impact of the unique UAV channel parameters on system performance, using the now-validated theoretical outage probability curves for a fixed FAS configuration. Fig.~\ref{fig:fig2a} shows the effect of multipath fading by varying $m$ while keeping $\alpha=2.1$ constant. As $m$ decreases from light fading, near-line-of-sight to severe fading, approaching Rayleigh-like conditions, the performance degrades greatly, as a lower $m$ implies a higher probability of deep fades. Fig.~\ref{fig:fig2b} shows the effect of shadowing by varying $\alpha$ while keeping $m=2.1$ constant. A more pronounced degradation is observed as $\alpha$ decreases from light shadowing to heavy shadowing. This demonstrates that severe shadowing, a key characteristic of A2G links, is a dominant factor in system performance, and validates that our model correctly captures this sensitivity.

\begin{figure}[!t]
\centering
\subfloat[Impact of $N$ ($W = 1.5\lambda$).]{
\includegraphics[width=0.47\columnwidth]{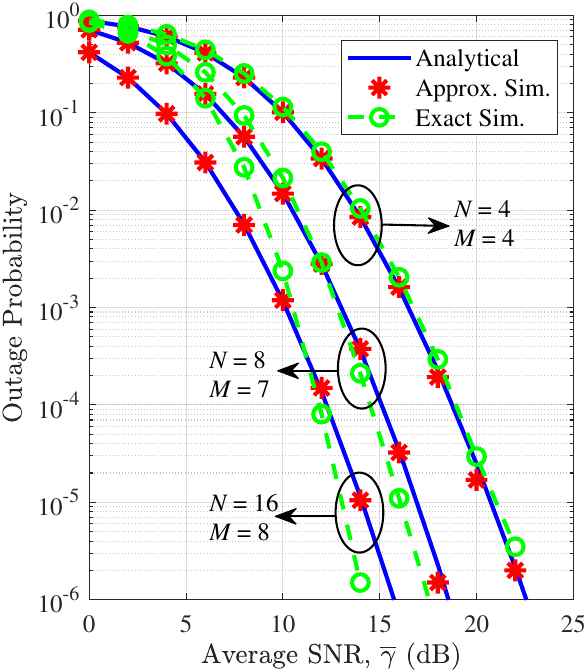}\label{fig:fig3a}%
}
\hfil
\subfloat[Impact of $W$ ($N = 8$).]{
\includegraphics[width=0.47\columnwidth]{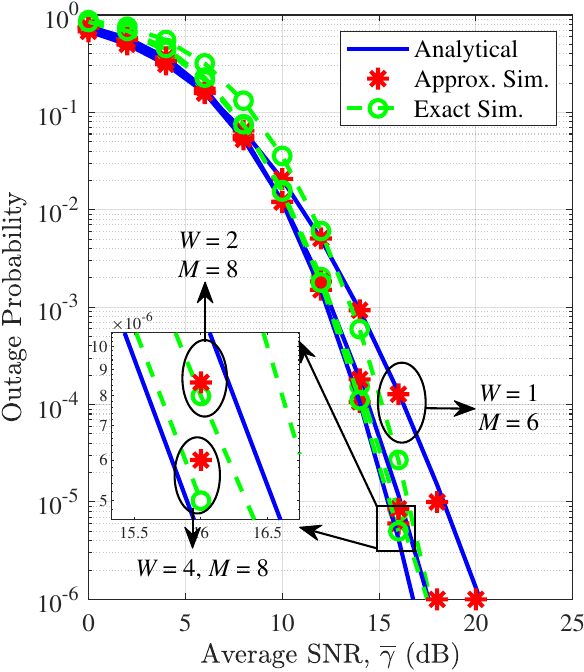}\label{fig:fig3b}%
}
\caption{Impact of FAS parameters on outage probability.}\label{fig:fig3_fas_params}
\vspace{-2mm}
\end{figure}

Fig.~\ref{fig:fig3_fas_params} serves a dual purpose, illustrating both the physical benefits of FAS and the fidelity boundaries of the eigenvalue-based analytical approximation. Fig.~\ref{fig:fig3a} clearly demonstrates rank saturation: for a fixed aperture ($W=1.5\lambda$), increasing the number of physical ports $N$ from $8$ ($M=7$) to $16$ ($M=8$) yields marginal diversity gain. This result confirms that the system's performance gain is fundamentally governed by the effective rank $M$, which quickly saturates within a small physical volume, regardless of further increases in $N$. Conversely, Fig.~\ref{fig:fig3b} highlights the decorrelation gain, where increasing the aperture $W$ allows $M$ to approach the physical limit ($M=N=8$), resulting in maximum performance. Critically, these results reveal the model's limitation: a measurable deviation exists between the Exact Simulation and the Approximate Simulation exclusively in the highly rank-deficient scenario ($N=16, M=8$). This divergence arises because the approximate model's predicted diversity order ($G_d = M \times d = 16.8$) theoretically exceeds the physical channel capacity constrained by the $N=16$ ports. Consequently, the approximation exhibits a predictable, optimistic bias when $G_d$ is constrained by $N$, confirming its highest fidelity is reserved for full-rank systems where $M=N$.

\section{Conclusion}\label{sec:conclusion}
In this letter, we established a novel analytical framework for an $N$-port FAS receiver operating over DS-UAV channels. By leveraging an eigenvalue-based approximation, we derived new analytical expressions for the end-to-end SNR statistics, e.g., both CDF and PDF, which enabled us to present exact integral definitions for the outage probability, ABER, and average channel capacity. We further presented new closed-form solutions for the ABER and capacity for the practical $M=2$ i.n.i.d.~case and proved that the system achieves a multiplicative diversity order of $G_d = M\times d$. This theoretical framework was comprehensively validated against Monte-Carlo simulations, which confirmed the high accuracy of the FAS approximation in Nakagami-$m$ channels. Our results also quantified the practical trade-offs between port density and physical aperture, confirming our model's applicability for realistic FAS design in UAV networks.

\end{document}